\documentclass{article}
\usepackage{emulateapj,apjfonts}
\usepackage{lscape}
\input psfig

\lefthead{Pseudobulge in NGC 4565}
\righthead{Kormendy \& Barentine}

\begin{document}

\centerline{\null}\vskip -15pt

\newcommand{\etal}{et~al.\ }

\title{Detection of a Pseudobulge Hidden Inside the ``Box-Shaped Bulge'' of NGC 4565}   
\author{John Kormendy and John C. Barentine}   
\affil{Department of Astronomy, University of Texas at Austin, 1 University
Station C1400, Austin, TX 78712-0259, USA}    

\begin{abstract} 
\noindent Numerical simulations show that box-shaped bulges of edge-on galaxies are not bulges: 
they are bars \hbox{seen side-on.}  Therefore the two components that are seen in edge-on Sb galaxies such 
as NGC 4565 are a disk and a bar.  But face-on SBb galaxies always show a disk, a bar, and a 
(pseudo)bulge.~Where is the (pseudo)bulge in NGC{\thinspace}4565?  We use archival {\it Hubble Space Telescope\/}
$H$-band images and {\it Spitzer Space Telescope\/} 3.6 $\mu$m wavelength images, both calibrated to 2MASS
$K_s$ band, to penetrate the prominent dust lane in NGC 4565.  We find a high surface brightness, central 
stellar component that is clearly distinct from the boxy bar and from the disk.  Its brightness profile is a S\'{e}rsic
function with index $n = 1.55 \pm 0.07$ along the major axis and 1.33$\pm$0.12 along the minor axis.  Therefore
it is a pseudobulge.  It  is much less luminous than the boxy bar, so the true pseudobulge-to-total luminosity 
ratio of the galaxy is  $PB/T = 0.06 \pm 0.01$, much less than the previously believed value of 
$B/T = 0.4$ for the ``boxy bulge''.  We infer that published $B/T$ luminosity ratios of edge-on
galaxies with boxy bulges have been overestimated.  Therefore, more galaxies than we thought
contain little or no evidence of a merger-built classical bulge.  From a formation point of view, NGC 4565 is a giant, 
pure-disk galaxy.  This presents a challenge to our picture of galaxy formation by hierarchical clustering: 
it is difficult to grow galaxies as big as NGC 4565 without also making big classical bulges. 
\end{abstract}


\begin{figure*}[b]

\includegraphics{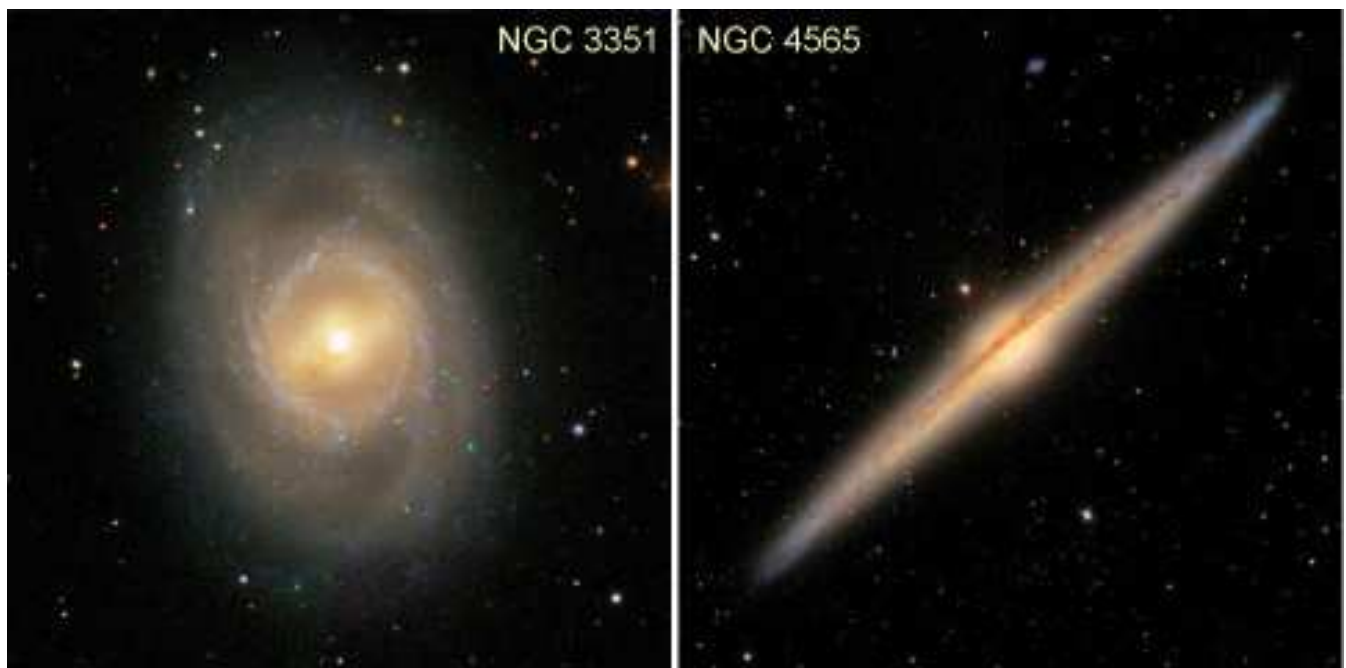}

\figcaption[]
{NGC 3351 (left) and NGC 4565 (right) in $gri$ composite color images from the Sloan Digital Sky Survey
(courtesy {\tt http://www.wikisky.org}).  Simien \& de Vaucouleurs (1986) estimate that $B/T = 0.1$ in
NGC 3351 (smaller than in NGC 4565) in part because they do not include the bar as part of the bulge.
}
\end{figure*}

\section{Introduction}

\pretolerance=15000  \tolerance=15000

      Figure 1 compares the prototypical Sb galaxies NGC 3351 and NGC 4565.  NGC 3351 is more nearly face-on
and shows three main components -- a bulge, a bar,~and~a~disk.  In contrast, NGC 4565 is 
edge-on; it shows only two components, a box-shaped bulge and disk.  As long as we thought that boxy structure 
was a secondary property of normal bulges, a galaxy morphologist (Sandage 1961) would just use the bulge-to-total
luminosity ratio $B/T \simeq 0.4$ (Simien \& de Vaucouleurs 1986) to classify NGC 4565 as an Sb.  However, 
we now know that \phantom{00000000}

\vskip 230pt

\centerline{\null} \vskip -6pt

\noindent   ``boxy bulges'' are not bulges at all; 
rather, they are edge-on bars (Combes \& Sanders 1981).  So the SBb galaxy NGC 3351 shows a disk, a bar, 
and a bulge, but the SBb galaxy NGC 4565 shows only a disk and a bar.  Where is the bulge in NGC 4565?

     Boxy bulges like that in NGC 4565 are a fundamental feature of edge-on disk galaxies 
(Sandage 1961; Buta \etal 2007).  Their identification as edge-on bars is a well known result.  
The Combes \& Sanders (1981) $N$-body model demonstration that bars heat themselves vertically 
by a combination of buckling instabilities and resonant star scattering has been confirmed
\phantom{00000000}

\clearpage

\noindent and extended many times
(Combes et al.~1990; 
Pfenniger~\& Norman 1990;
Pfenniger \& Friedli 1991; 
Raha et al.~1991;
Athanassoula \& Misiriotis 2002; 
Athanassoula 2005, 
Shen et al.\ 2010, and many others).
Cylindrical rotation is observed in $N$-body bars and in boxy bulges
(Kormendy \& Illingworth 1982;
Jarvis 1990;
Shaw, Wilkinson, \& Carter 1993;
Bettoni \& Galletta 1994;
Fisher \etal 1994;                    
D'Onofrio et al.~1999;                
Falc\'on-Barroso et al.~2004;         
Howard \etal 2008)         
but not in classical, elliptical-galaxy-like bulges 
(Illingworth \&  Schechter 1982; 
Kormendy \& Illingworth 1982; 
Binney \etal 1990;                        
Verolme et al.~2002;               
Copin \etal 2004;                     
Emsellem \etal 2004);                
this further cements the connection between boxy bulges and edge-on bars.  
Finally, a splitting of gas rotation velocities in edge-on boxy bulges (a ``figure 8'' shape
of spectral emission lines) also is a robust signature of gas flow in an edge-on bar
(Kuijken \& Merrifield 1995;
Merrifield 1996;
Merrifield \& Kuijken 1999;
Bureau \& Freeman 1999).
There is little doubt that the ``boxy bulge'' of NGC 4565 is not the real bulge of the
galaxy.  Does the galaxy contain a bulge at all?  That is, does it contain a dense,
central component that we would identify as a bulge
in addition to the bar if the galaxy were seen face-on?

      We care for two reasons, both involving galaxy formation:

      {\it Background:} Early galaxy evolution was dominated by hierarchical gravitational clustering of density 
fluctuations that resulted in galaxy collisions and mergers \hbox{(White \& Rees 1978);} these scrambled disks
into ellipticals (Toomre~1977).~Enormous energy has been invested in studying hierarchical clustering;
there is little danger that the picture is fundamentally wrong (Binney 2004).  However, it is incomplete. 
Recent work has established that hierarchical clustering is gradually giving way to a complementary suite 
of evolution processes that shape {\it isolated\/} galaxies.  They evolve by rearranging energy and angular 
momentum; one consequence is the growth of central components that masquerade as classical bulges but that, 
in general, formed slowly (``secularly'') out of disks (see Kormendy 1993; Kormendy \& Kennicutt 2004
for reviews).  We call them ``pseudobulges'' to distinguish them from merger remnants.  They come in 
at least two varieties.   As reviewed above, ``boxy bulges'' are believed to be edge-on bars.  
Our Galaxy contains one (Dwek \etal 1995).   Another variety is grown out of 
disk gas that was transported inward by nonaxisymmetries such as bars; we call them ``disky pseudobulges'' 
here, because they are often highly flattened, but we emphasize that they are not always flat (Kormendy 1993;
Kormendy \& Kennicutt 2004; \S\thinspace2 here).  Then:

      {\it Reason 1:} Confidence in our conclusion that the boxy center of NGC 4565 is an edge-on bar
would be increased if we also observed a (pseudo)bulge as we do in face-on galaxies~(Fig.~1).  
As long as face-on and edge-on galaxies appear to show physical differences,
we can't be sure that we understand them.

      {\it Reason 2:} If the box in NGC 4565 is a bar, then it is part of the disk and $B/T$ is smaller than 
we thought.  If in addition the galaxy contains a pseudobulge and not a hidden classical bulge, 
then $B/T$ is even smaller -- possibly zero.  This is hard to understand in the context of hierarchical 
clustering, which essentially always makes substantial bulges in giant galaxies (see, e.{\thinspace}g.,
Peebles \& Nusser 2010; Kormendy \etal 2010).

\section{Dissecting NGC 4565 Using Mid-Infrared Images}

      Fundamental plane correlations tell us that low-luminosity bulges are small and dense, not
large and fluffy 
(Djorgovski \& Davis 1987; 
Faber \etal 1987;
Bender \etal 1992;
Kormendy \etal 2009).
So a (pseudo)bulge in NGC 4565 could only hide behind the dust lane.  We use {\it Spitzer Space Telescope} IRAC 
archive images at 3.6 $\mu$m and 8 $\mu$m to look through the 
dust and measure the light profile of NGC 4565.  The images are shown in Figure 2,  The contrast and brightness 
(``stretch'') in the top panel emphasize the boxy pseudobulge.  A lighter stretch in panel (b) 
reveals two new features that are hidden at optical wavelengths.  First, the galaxy shows an ``inner ring'' 
like those associated with bars (Sandage 1961; Buta \etal 2007).  The 
3.6 $\mu$m image shows starlight at the reddest wavelength that is free from dust emission.  This
wavelength is long enough so that dust absorption is small.  Still, there is a small danger that the ring
looks dark inside because of absorption.  But the 8 $\mu$m image in panel (c) shows PAH emission
from dust.  The ring is bright, implying active star formation.  But there is no emission from
inside the ring except at the center.  We conclude that NGC 4565 is an almost-edge-on SB(r) galaxy.
It is similar to NGC 2523, which is shown in Figure 2(d) scaled to match the ring
size in NGC 4565 and with an orientation similar to the one that we infer for NGC 4565, with the bar almost
along the line of sight.  Our understanding of vertical bar thickening implies that NGC 2523, oriented
as in Figure 2(d) but seen more nearly edge-on, would resemble NGC 4565.

      The second new feature shown in Figure 2(b) proves to be the ``missing'' pseudobulge.  It is the tiny bright
region at the galaxy center.  It is compact -- as expected -- and it appears clearly distinct from the much 
lower surface brightness bar.  To check whether it really is distinct from the bar and to see whether it is
a small classical bulge or a pseudobulge, we measured its surface brightness profile along the major and
minor axes of the galaxy.  The minor-axis profile is shown in Figure 3.  {\it Spitzer\/} resolution is 
poor, so we supplemented the {\it Spitzer\/} profile at small radii by measuring a
{\it Hubble Space Telescope\/} (HST) NICMOS F160W archive image.  Brightness cuts were extracted 
along the major and minor axes of NGC 4565.  We used the minor-axis profile only on the side of the
galaxy that is less affected by absorption.  All profiles were zeropointed to the $K_s$ bandpass of the
2MASS Large Galaxy Atlas (Jarrett \etal 2003).

      The minor-axis brightness profile is plotted in Fig.~3
against (radius)$^{1/4}$ so the profile of a classical bulge is nearly a straight line.  The profile
consists of three separate segments, an outer halo, an \hbox{intermediate-$r$} profile
that describes the boxy structure (see Kormendy \& Bruzual 1978), and the dense central component.  The
latter components have concave-downward profiles, so they are S\'ersic (1968) functions
with indices $n \ll 4$.  We therefore made a three-component decomposition into an outer exponential and two
S\'ersic functions.  The intermediate-radius profile of the boxy bar proves to be exponential, $n = 1$.  The
central profile has $n = 1.33 \pm 0.12$ along the minor axis and $n = 1.55 \pm 0.07$ along the major axis.  
Both values are robustly less than 2.  Classical bulges have $n \gtrsim 2$ whereas most pseudobulges have 
$n \lesssim 2$ (e.{\thinspace}g., Fisher \& Drory 2008).   Thus both the boxy structure and the inner bright 
region are pseudobulges, and NGC 4565 contains both subtypes of pseudobulge.  However, the disky 
pseudobulge -- the central component -- is not flat: it has an axial ratio of $\sim 0.8$ to 0.9.

      Remarkably, the smallest scale height of any structural component in the galaxy is that of the pseudo{\it bulge}.
The decomposition in Figure 3 gives its exponential scale height as $\sim 1\farcs2 \simeq 90$ pc
(we adopt a distance to NGC 4565 of 14.5 Mpc from Wu \etal 2002).  We measure an exponential scale height 
of boxy bar plus disk of $10\farcs5 \simeq 0.74$ kpc.  The scale heights of the thin and thick disks have been 
measured by many authors (e.{\thinspace}g., van der Kruit \& Searle 1981, Jensen \& Thuan 1982, Shaw

\centerline{\null} 

\begin{figure*}[b]

\includegraphics{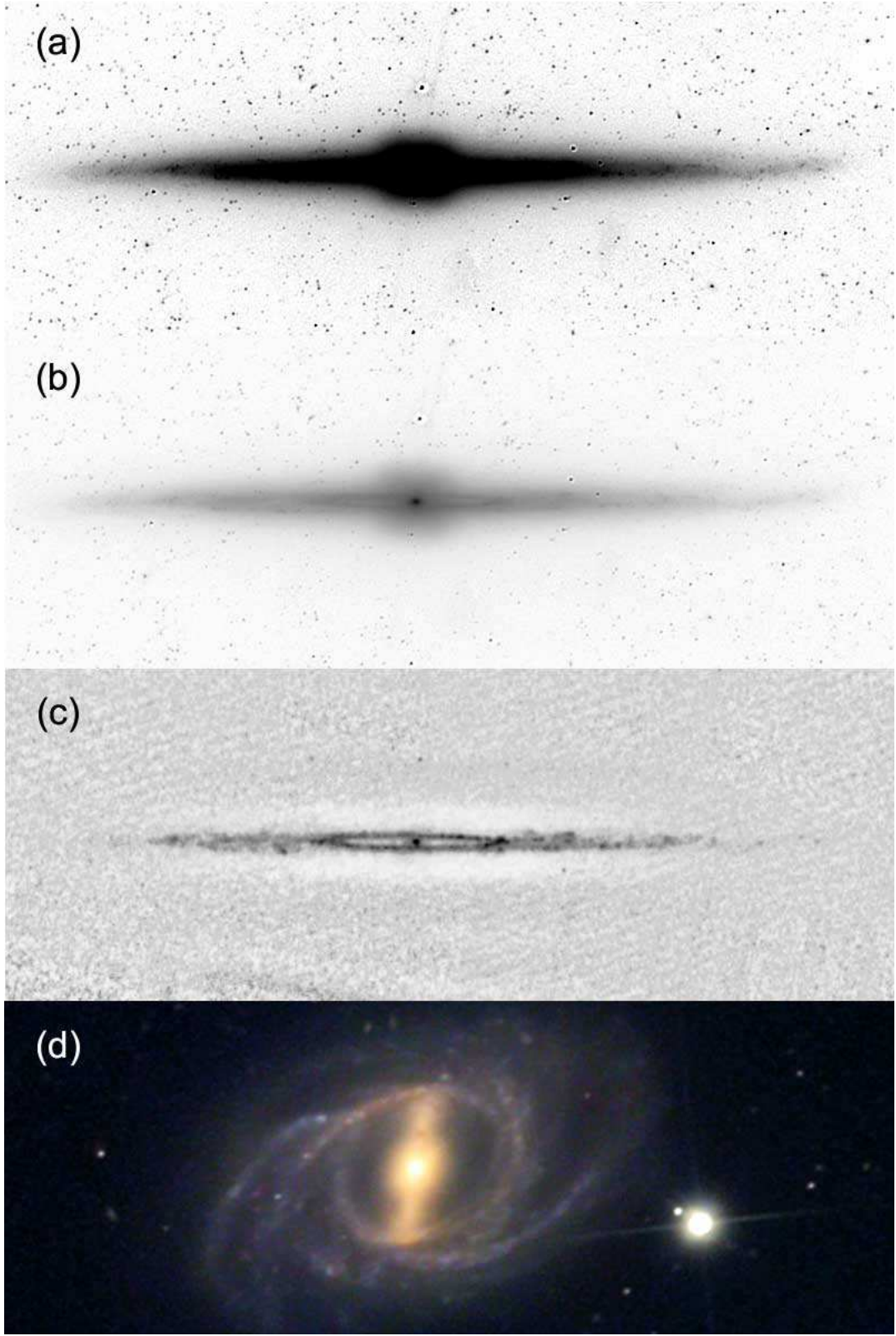}

\figcaption[]
{(a and b) PSF-deconvolved Spitzer/IRAC 3.6 $\mu$m negative images of NGC 4565 shown at different stretches
 that emphasize (a) the boxy bar and (b) an inner ring and pseudobulge.  The newly 
detected pseudobulge is the tiny bright spot at the galaxy center.  Its scale height is smaller than 
that of the outer disk.  (c) Spitzer/IRAC 8 $\mu$m negative image showing PAH emission and 
therefore star formation from the inner ring and outer disk.   Active star formation is normal in inner
rings (Kormendy \& Kennicutt 2004).  The observation that the inner ring is 
dark inside at 8 $\mu$m means that the apparently dark inside seen at 3.6 $\mu$m is not caused by dust 
absorption.  Rather, the ring really is dark inside.  We conclude that NGC 4565 is an SB(r)b galaxy;
that is, an almost-edge-on  analog of the SB(r)b galaxy NGC 2523 (bottom panel, in a positive, true-color 
image).  The NGC 2523 image has been scaled so the inner ring has the same apparent radius 
as in NGC 4565 and rotated to approximately the apparent bar position angle inferred for NGC 4565.  
That is, we suggest that the bar of NGC 4565 is seen almost end-on and that, if NGC 2523, oriented as in the
bottom panel, were inclined still more until we observed it almost edge-on, it would show the features 
seen in the NGC 4565 images.  To put it another way, we suggest that, if NGC 4565 were observed face-on,
it would be the most spectacular SB(r) galaxy in the nearby Universe.
}
\end{figure*}

\eject

\centerline{\null}

\vskip 2.6truein

\includegraphics{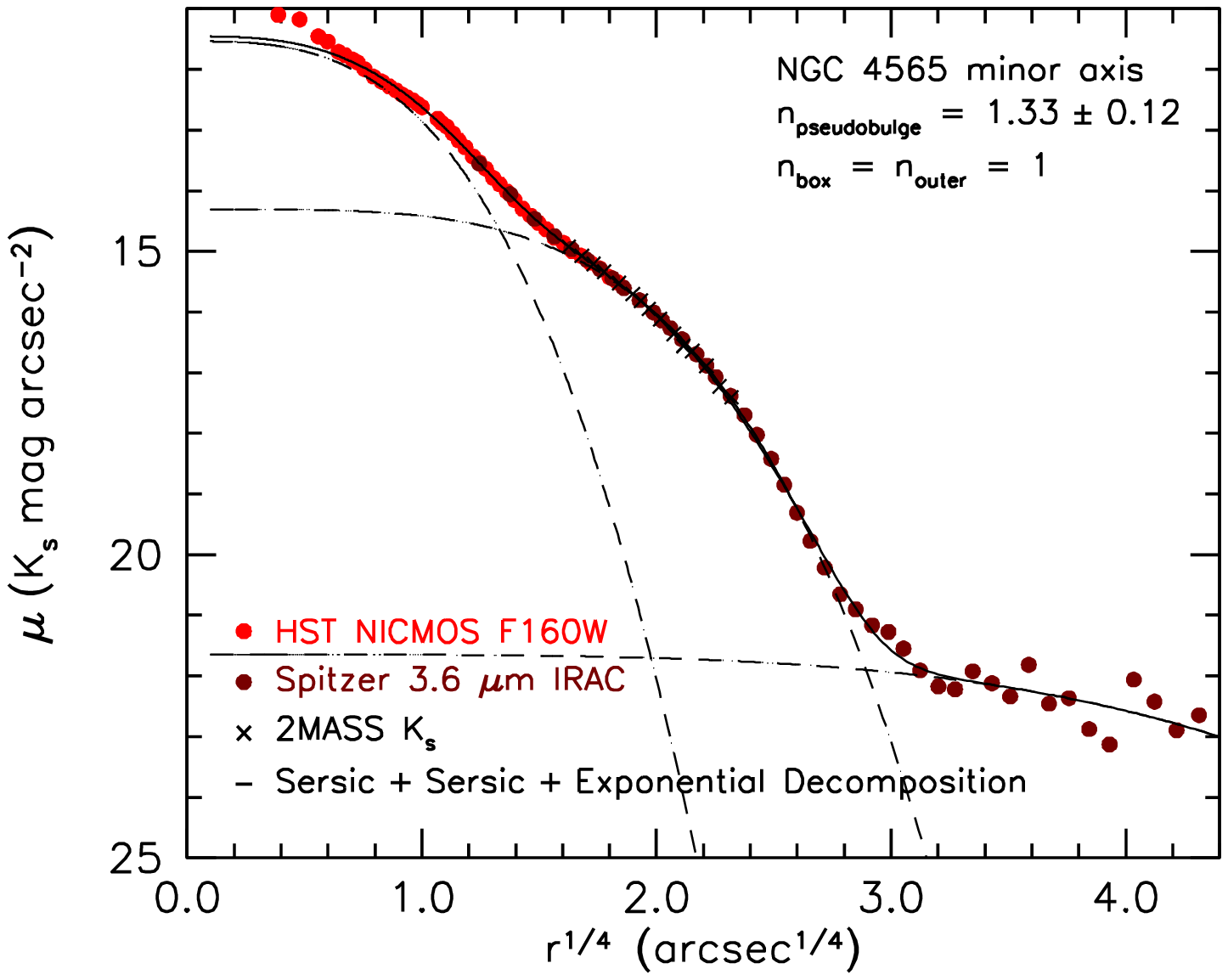}

\figcaption[]
{Minor-axis profile of NGC 4565 from HST NICMOS F160W (red points) and Spitzer IRAC 3.6 $\mu$m 
(brown points).  A profile (black crosses) calculated from the 2MASS Large Galaxy Atlas image 
(Jarrett \etal 2003) provides the $K_s$ zeropoint.  Dashed lines show a decomposition 
of the profile into components in order of increasing radius: pseudobulge (S\'{e}rsic), 
box-shaped bar (S\'{e}rsic), and outer halo (exponential, but our data do not constrain the
functional form).  The solid line is the sum of the components.  A central Seyfert nucleus (Ho \etal 1997) 
or star cluster is not included in the fit,
}

\vskip 5pt

\noindent  \& Gilmore 1989, and
Wu \etal 2002, who also tabulate previous results).   Seven measurements of the thick disk scale height average
to $14\farcs6 = 1.03$ kpc with a dispersion of $2\farcs3$.  Eight measurements of the scale height of the thin 
disk average $8\farcs0 = 0.56$ kpc with a dispersion of $0\farcs9$.  The middle component
in Figure 3 has a scale height intermediate between that of the thin and thick disk, as expected, given that
it includes thin disk, thick disk and boxy bulge along the line of sight.  But the pseudobulge has a scale 
height that is smaller than that of the thin disk.

      Simien \& de Vaucouleurs (1986) find a bulge-to-total light ratio of $B/T$ = 0.4.  However,
$B$ refers to the boxy bar -- not the pseudobulge within.  Figure 2 shows that the pseudobulge is much less 
luminous than the boxy structure.  If NGC 4565 were seen face-on, what we identify as a box-shaped bulge 
would be recognized as a bar and would not be included in the bulge light inventory.  Previously measured $B/T$ 
ratios of edge-on galaxies with box-shaped bulges are therefore overestimated.

      For the central pseudobulge, we measure a $K_s$ magnitude of $9.09 \pm 0.15$.
Jarrett \etal (2003) find that the total $K_s$ magnitude of the galaxy is $6.060 \pm 0.017$.  So
the (inner) pseudobulge-to-total luminosity ratio is $PB/T = 0.06 \pm 0.01$, similar to $B/T$ 
in NGC 3351 (Figure 1).  We see no sign of a classical bulge.  

      Indeed, a normal classical bulge with $B/T \sim 0.06$ in $V$ band would have an absolute magnitude of 
$M_{V,{\rm bulge}} \simeq -18.4$ and an effective radius of $r_e \simeq 0.7$ kpc = 10$^{\prime\prime}$ 
(Kormendy \etal 2009).  It would be very obvious.  
We conclude that $B/T \ll 0.06$. In other words, NGC 4565 is effectively bulgeless.
It is very similar to our own Galaxy (e.{\thinspace}g., Howard \etal 2008; Shen \etal 2010).

\section{Conclusion}

      The interpretation of boxy bulges in edge-on Sb galaxies as bars is more believable if we also 
find (pseudo)bulges like those associated with bars in face-on Sb galaxies.  Our discovery of a pseudobulge
in NGC 4565 that is distinct from the boxy bar increases confidence in our picture of secular evolution.   

      Furthermore, $B/T$ ratios in edge-on galaxies with boxy bulges are smaller than previously 
believed.   In NGC 4565, the detection only of a central pseudobulge means that we see no sign of a 
major merger remnant.   Moreover, NGC 4565 rotates at 255 km s$^{-1}$ interior 
to the outer warp (Rupen 1991).   NGC{\thinspace}4565 has grown very massive while remaining 
a pure-disk galaxy.  We do not know how this can happen in a Universe dominated 
by the dynamical violence of hierarchical clustering.

\centerline{\null} \vskip -30pt \centerline{\null}

\acknowledgements 

These results are based on observations made with the {\it  Spitzer Space Telescope}, which is
operated by JPL under contract~with NASA.  Additional data from the NASA/ESA {\it Hubble Space Telescope} 
were obtained from the data archive at the Space Telescope Science Institute.  It is operated by the AURA, 
Inc., under NASA contract NAS 5-26555.  This work was supported by the National Science Foundation 
under grant AST-0607490. 

\centerline{\null}\vskip -25pt
\centerline{\null}

\end{document}